# Anisotropy of Antiferromagnetic Domains in a Spin-orbit Mott Insulator

Longlong Wu[1,*], Wei Wang[1], Tadesse A. Assefa[1,2], Ana F. Suzana[1], Jiecheng Diao[3], Hengdi Zhao[4], Gang Cao[4], Ross J. Harder[5], Wonsuk Cha[5], Kim Kisslinger[6], Mark P. M. Dean[1] and Ian K. Robinson[1,3†]

[1]*Condensed Matter Physics and Materials Science Department, Brookhaven National Laboratory, Upton, NY 11973, USA*

[2]*Stanford Institute for Materials and Energy Sciences, SLAC National Accelerator Laboratory, Menlo Park, California 94025, USA*

[3]*London Centre for Nanotechnology, University College London, London, WC1E 6BT, United Kingdom.*

[4]*Department of Physics, University of Colorado at Boulder, Boulder, Colorado 80309, USA*

[5]*Advanced Photon Source, Argonne National Laboratory, Illinois 60439, USA*

[6]*Center for Functional Nanomaterials, Brookhaven National Laboratory, Upton, New York 11793, USA*

## Abstract

The temperature-dependent behavior of magnetic domains plays an essential role in the magnetic properties of materials, leading to widespread applications. However, experimental methods to access the three-dimensional (3D) magnetic domain structures are very limited, especially for antiferromagnets. Over the past decades, the spin-orbit Mott insulator iridate $Sr_2IrO_4$ has attracted particular attention because of its interesting magnetic structure and analogy to superconducting cuprates. Here, we apply resonant x-ray magnetic Bragg coherent diffraction imaging to track the real-space 3D evolution of antiferromagnetic ordering inside a $Sr_2IrO_4$ single crystal as a function of temperature, finding that the antiferromagnetic domain shows anisotropic changes. The anisotropy of the domain shape reveals the underlying anisotropy of the antiferromagnetic coupling strength within $Sr_2IrO_4$. These results demonstrate the high potential significance of 3D domain imaging in magnetism research.

**Introduction**

In magnetic materials, the formation and dynamics of the magnetic domains play a critical role to determine the macroscopic magnetic properties, such as magnetization and magnetic susceptibility [1-3]. Experimental access to mesoscale structures of magnetic materials can provide information about fundamental magnetic interactions [4-8], which may be related to intrinsic topological properties in the quantum matter [4]. An intriguing Ruddlesden-Popper family of layered two-dimensional (2D) antiferromagnetic spin-orbit Mott insulators, the iridate series $Sr_{n+1}Ir_nO_{3n+1}$, presents an interesting class of frustrated systems, where the spin, orbital, and crystal field interactions conspire to determine the overall electronic configuration [9,10]. Indeed, the single-layer iridate, $Sr_2IrO_4$, which we study here, already hosts a variety of unconventional magnetism [11]. In $Sr_2IrO_4$, the narrow electronic band of spin-orbit-split Ir $5d$ states can be further split by the modest on-site Coulomb repulsion to generate an antiferromagnetic Mott insulating state [10,12-14].

The antiferromagnetic Mott ground state of $Sr_2IrO_4$ can be tuned by different parameters, for example, chemical composition, temperature, strain, and disorder [15,16]. Understanding how magnetic domains respond to the presence of disorder can be detected in their robustness to temperature variations. The structure of $Sr_2IrO_4$ has been reported to be tetragonal, yet the antiferromagnetic structure breaks this symmetry in the ab-plane which should lead to orthorhombic domains discussed below. However, x-ray magnetic scattering experiments [17] report the formation of a single magnetic domain orientation in crystals of millimeter size. This is probably due to the presence of some residual strain every time the crystal is cooled in the cryostat. Only in the large samples needed for neutron scattering do the crystals adopt an equal mixture of domain orientations [18]. The antiferromagnetic domain sizes reported in these experiments are all resolution-limited, showing that the antiferromagnetic ordering is strong. As a notoriously difficult task, imaging antiferromagnetic structure is accessible to some existing microscopy techniques, for example, x-ray photoemission electron microscopy [19], spin-polarized scanning tunneling microscopy [20], propagation-based phase contrast technique [21], Bragg ptychography [22], and optical second-harmonic generation technique [23,24]. They mainly provide two-dimensional structure information of magnetic materials. Bragg Coherent Diffractive Imaging (BCDI) is a powerful x-ray technique for imaging nanoscale structures in three dimensions [25,26]. It requires a coherent x-ray beam, available from the undulators of third and fourth generation synchrotron sources, such as the Advanced Photon Source (APS). But even there, a large fraction of the available x-ray flux is lost to preserve the beam coherence and so, until now, it has not been possible to overcome the small resonant magnetic cross-section to image antiferromagnetic materials.

In this work, we use the resonant x-ray magnetic BCDI approach to study the temperature-dependence evolution of the three-dimensional (3D) image of the antiferromagnetic domains within a $Sr_2IrO_4$ single crystal. The small $Sr_2IrO_4$ crystal was fabricated by Focused Ion Beam (FIB) milling. Inverted Bragg coherent diffraction images from the crystal show the formation of antiferromagnetic domains when the temperature of the $Sr_2IrO_4$ crystal is below the Néel transition temperature $T_N$~230K. Additionally, from reconstructed images, the estimated antiferromagnetic domain size is ~1.1 μm at 120 K and therefore fills the host crystal in the in-plane directions. Upon increasing the temperature, the antiferromagnetic domain is seen to shrink along the c-axis direction, suggesting the anisotropic coupling between the Ir spins.

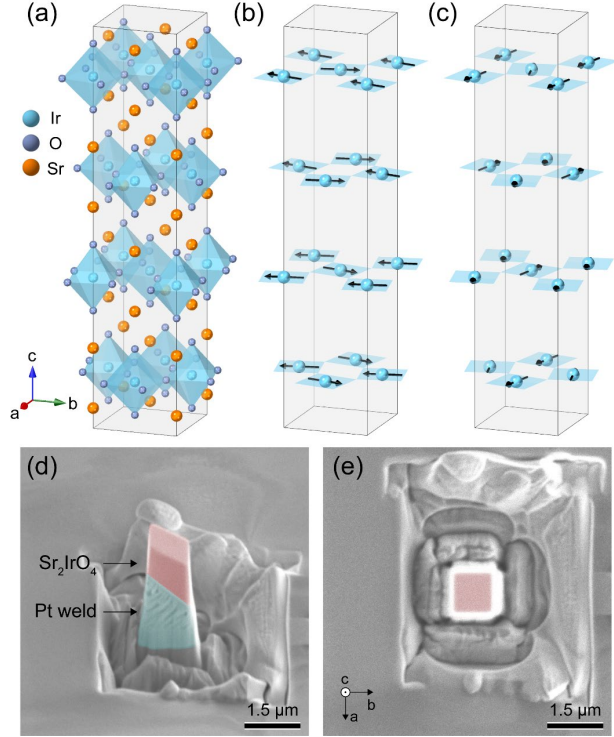

FIG. 1. (a) The crystal structure of $Sr_2IrO_4$ with space group $I4_1/acd$. (b) $ab$ plane canted antiferromagnetic structure. The canting angle of the magnetic moment follows the octahedral rotation rigidly. (c) Corresponding twin domain states of the antiferromagnetic structure. (d) Scanning electron microscopy image of the $Sr_2IrO_4$ sample from the side view. (e) Top view. The corresponding crystal orientation is indicated in (e).

$Sr_2IrO_4$ has the classic "214" layered perovskite structural motif with the family of cuprate superconductors. Its lattice constants are $a = b = 5.49$ Å and $c = 25.80$ Å at room temperature with tetragonal symmetry (space group $I4_1/acd$), as shown in Fig. 1(a). In $Sr_2IrO_4$, the square lattice of $Ir^{4+}$ ions are formed by corner-shared $IrO_6$ octahedra, which elongate and rotate about the c-axis by $\phi = \sim 11.8°$ [18,27]. Below the Néel temperature $T_N \sim 230$ K, the spin-orbital entangled $J_{eff} = 1/2$ magnetic dipole moments undergo 3D long-range ordering into an orthorhombic antiferromagnetic structure that preserves global inversion symmetry but has lower rotational symmetry. Figure 1(b) and (c) show the two canted antiferromagnetic structures on the $IrO_2$ planes of $Sr_2IrO_4$, with the black arrows indicating pseudospins and cyan squares indicating the $IrO_2$ planes [11,28]. As presented, when the temperature is below $T_N$, $Sr_2IrO_4$ has two possible canted antiferromagnetic structures with the net spin moment along the a- and b-axes, separately [11,29]. These antiferromagnetic structures are related to the $4_1$ screw of the $Sr_2IrO_4$ lattice structure. As a result, the model of twin antiferromagnetic domain states I and II [i.e., with antiferromagnetic structures described by Fig. 1(b) and (c), separately] in $Sr_2IrO_4$ generate magnetic reflections at $\mathbf{Q}_I$ = (1, 0, 4n + 2) and (0, 1, 4n) [28], while for domain state II, the magnetic reflections are presented only at $\mathbf{Q}_{II}$ = (1, 0, 4n) and (0, 1, 4n + 2).

The high-quality $Sr_2IrO_4$ single crystal was grown from off-stoichiometric quantities of $SrCl_2$, $SrCO_3$, and $IrO_2$ using self-flux techniques [29]. To obtain the micro-size crystal needed for the BCDI experiment [30], a large $Sr_2IrO_4$ crystal was firstly pre-oriented crystallographically using a Laue diffractometer, then a block with its size around 1.2 µm was cut out using the FIB lift-out method (Supplemental Material [31]). Afterward, the obtained $Sr_2IrO_4$ crystal was welded with Pt onto a silicon wafer as shown in Fig. 1(d)-(e). The selected size of the crystal is less than the penetration depth, ~6.6 $\mu m$ (Supplemental Material) of 11.218 keV x-ray so that the dynamical diffraction effect is minimized.

The resonant x-ray magnetic BCDI experiment was carried out at the 34-ID-C beamline of the APS, Argonne National Laboratory (Supplemental Material [31]). The incident x-ray energy was tuned to the peak of the magnetic signal at 11.218 keV with $\pi$ polarization, just below Ir $L_3$ edge (i.e., the transition from $2p_{3/2}$ to $5d$ state) of the $Sr_2IrO_4$ crystal. During the measurements, the crystal was mounted on a micro miniature refrigerator (MMR) stage where the temperature was controlled by an MMR K-20 controller. The MMR technology uses a heat exchanger and Joule-Thompson expansion under a fixed flow of high-pressure nitrogen gas [38,39]. Since the $Sr_2IrO_4$ crystal was pre-aligned before the FIB preparation, the crystal alignment was quickly determined using the crystal (204) and (116) peaks with a six-axis diffractometer to maneuver its orientation.

The coherent diffraction data were collected by rocking the sample. The crystal (116) and antiferromagnetic (106) peaks were used to study the crystal and magnetic structures, separately.

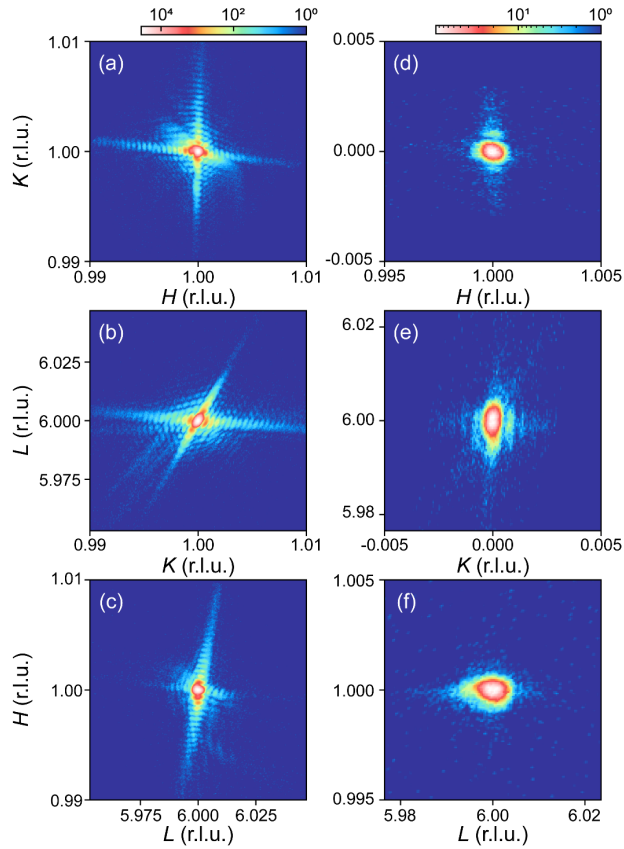

FIG. 2. (a)-(c) Central slice of the 3D reciprocal space mapping around the (116) Bragg peak. (d)-(f) Central slice of the 3D reciprocal space mapping around the antiferromagnetic (106) peak, which was obtained at T=120 K.

Figure 2 shows representative Bragg coherent diffraction patterns of the crystal (116) and antiferromagnetic (106) peaks from the $Sr_2IrO_4$ crystal, converted into reciprocal lattice unit (r.l.u.) coordinates from the detector coordinates. Here, the reciprocal space is indexed with the tetragonal unit cell, defined as $\mathbf{Q} = \left(\frac{2\pi}{a}H, \frac{2\pi}{b}K, \frac{2\pi}{c}L\right)$. As presented in Fig. 2(a)-(f), these two Bragg coherent x-ray diffraction patterns have different diffraction fringe distributions but similar directions, indicating their distinct sizes and similar facet orientations in real space between the crystal and antiferromagnetic structures. When the crystal is perfect, for the (116) peak, the coherent x-ray diffraction pattern is mainly determined by the external shape of the crystal. Since the crystal domain size sets an upper limit for the magnetic domain size, the fringe spacings from the antiferromagnetic (106) peak should be similar to or larger than those of the crystal (116) peak.

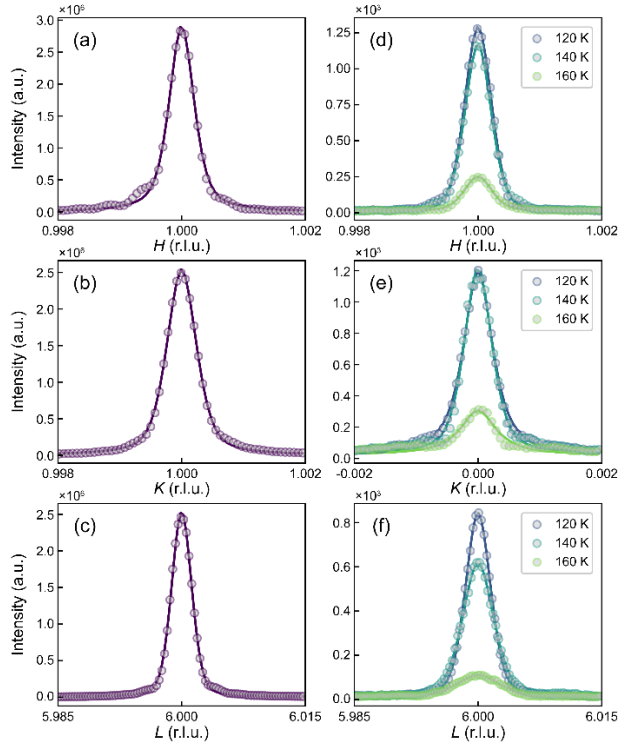

FIG. 3. (a)-(c) Integrated scan intensity for the wave-vector scan along (a) the [$H$, 0, 0], (b) the [0, $K$, 0], and (c) the [0, 0, $L$] directions from the crystal (116) peak. (d)-(f) Corresponding integrated scan intensity from the antiferromagnetic (106) peak at the selected temperatures along different directions.

To explore the effect of temperature on the antiferromagnetic domain in the Sr$_2$IrO$_4$ crystal, Fig. 3 shows the integrated x-ray intensity along the [$H$, 0, 0], [0, $K$, 0] and [0, 0, $L$] direction separately from the (106) peak at different selected temperatures and the accompanying (116) peak. Each scan is integrated along the other two transverse r.l.u. directions. As shown in Fig. 3(d)-(f), the antiferromagnetic peak is getting broader along the [0, 0, $L$] direction as the temperature is increasing, indicating the antiferromagnetic domain is shrinking along the c-axis direction. However, changes of the peak width in the other two directions are not apparent. To quantitatively determine the effect of the temperature on the antiferromagnetic domain, these one-dimensional lineshapes were fitted with Pseudo-Voigt functions. Table I shows the corresponding temperature dependence of the correlation length along the different directions, defined as $\xi = d/\text{FWHM}$, where FWHM represents for full width at half maximum in reciprocal lattice units and $d$ is the lattice spacing of the appropriate reflection.

TABLE I. Correlation lengths of the antiferromagnetic domain along different directions at the selected temperature. Here, the second column shows the correlation length of the host crystal itself by using the crystal (116) peak.

| Correlation Length [nm] | (116) peak | 120 K | 140 K | 160 K |
|---|---|---|---|---|
| $\xi_H$ | 1153± 6 | 1137± 4 | 1158± 4 | 1030 ± 13 |
| $\xi_K$ | 970 ± 3 | 1019± 15 | 1083± 20 | 871± 54 |
| $\xi_L$ | 902± 3 | 739 ± 2 | 610 ± 3 | 423± 9 |

We consider that the slightly larger value of the correlation length from the antiferromagnetic (106) peak along the [0, K, 0] direction at 120 and 140 K than that of the crystal (116) peak, seen in Table I, is due to a slight rotation of the $Sr_2IrO_4$ crystal under illumination with the highly coherent beam, opposite to the scanning direction, as could also be seen from one scan to the next [40]. However, the effect of crystal movement along the other directions, which are mainly parallel to the detector face, is small. Nevertheless, when the temperature is increased to 160 K, the correlation lengths of the antiferromagnetic domain are decreased along all directions by a different amount. This trend indicates a c-axis to a/b-axis anisotropy of the antiferromagnetic coupling strength along different crystal directions. Specifically, the coupling of spins within the $IrO_2$ a/b plane is indicated to be stronger than between spins between neighboring planes along the c-axis.

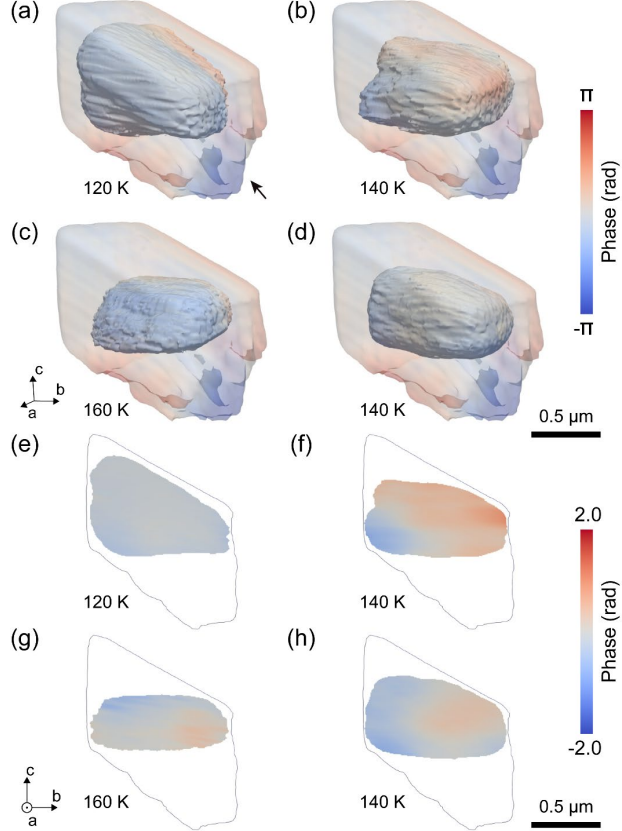

FIG. 4. (a)-(c) Antiferromagnetic domain obtained at T=120, 140, and 160 K, respectively. (d) Antiferromagnetic domain at T=140 K after the thermal cycling. Translucent images of the crystal from the (116) peak and isosurface images from the antiferromagnetic (106) peak are superimposed. (e)-(h) Corresponding center slice of (a)-(c). Here, the black lines show the profile of the $Sr_2IrO_4$ crystal. The antiferromagnetic images were contoured at 25% of the maximum amplitude and colored according to the image phase at the surface.

Compared with the reciprocal-space data fitting method, real-space images can give different details about the temperature-dependent evolution of the antiferromagnetic domain. Since the data in Fig. 2 and 3 were measured with a coherent x-ray beam, the measured coherent x-ray diffraction patterns can be inverted using the phase retrieval method [35,36]. Figure 4(a)-(c) show the corresponding reconstructed results using the coherent diffraction data from Fig. 2 and 3. Meanwhile, Figure 4(e)-(h) present the corresponding central slice of Fig. 4(a)-(c). As presented in Fig. 4, the translucent isosurface is obtained from the (116) peak at room temperature, while the inside shows the corresponding antiferromagnetic domain at the different temperatures. Compared with the SEM image in Fig. 1, the missing part below the bottom corner of the reconstruction of the Sr$_2$IrO$_4$ crystal, as indicated by the black arrow in Fig. 4(a), is probably due to the sample damage induced by the FIB milling process [41,42]. Also, this region of the crystal is found to contain significant strain, indicated by strong phase values in the complex image, interpreted as a projection of the lattice displacement onto the (116) measurement **Q** vector [25]. This and similar behavior confirmed from another Sr$_2$IrO$_4$ crystal (~1.2 μm in size) prepared by the same FIB lift-out method are shown in the Supplemental Material [31]. As the temperature is increasing from 120 K to 160 K, the antiferromagnetic domain shows an anisotropic change. Especially starting from the top corner of the crystal along the c-axis direction, the size of the antiferromagnetic domain decreases much more than other directions. Additionally, the observed antiferromagnetic domains do not occupy the entire volume of the Sr$_2$IrO$_4$ crystal, as shown in Fig. 4(a), most notably in the c-axis direction, also seen in Table I. It is possible that these empty regions are filled with the other antiferromagnetic domain orientation, which was not possible to be detected using the (016) peak, due to the limited in-plane rotation angle range of the MMR stage. Therefore, we tried to measure the antiferromagnetic (108) peak of this crystal but could not find any magnetic signal, indicating in these regions the domains with antiferromagnetic spin components along the other direction are too small to be detected or they are not existing. On a larger Sr$_2$IrO$_4$ sample (~8 μm in size), it was checked that when the (106) peak appears, the (108) peak of the other domain also appears (Supplemental Material [31]). Thus, as shown in Fig. 4(a), the disappearance of the antiferromagnetic domain in the empty region is possibly related to the strong residual strain induced by the FIB milling process, where the antiferromagnetic domains become orientationally pinned along the b-axis, which cannot be reached. Another possibility is that the strong strain induced by the FIB milling and Pt-welding process suppresses the formation of the antiferromagnetic domain as a large strain is formed at the bottom corner of the $Sr_2IrO_4$ crystal. Meanwhile, a weak strain is also observed inside the antiferromagnetic domains, as shown in Fig. 4. While the single antiferromagnetic domain almost fills the crystal in the a- and b-axis directions, there is a negligible morphological change for the domain near the bottom corner of the crystal when changing the temperature from 120K to 160K. Away from the bottom corner of the $Sr_2IrO_4$ crystal, the shortening of the domain size starting from the top corner of the crystal along the c-axis direction is most probably due to the relatively weaker out-of-plane spin-spin interaction compared to the in-plane interaction and intrinsic to the iridate compound but may also be caused by the interaction from the small domains at the top corner if existing.

It is hard to know the exact locations of the antiferromagnetic domain in the host $Sr_2IrO_4$ crystal in Fig. 4, since the two images come from independent measurements. In the a- and b-axis directions, the location is strongly constrained by geometry. But in the c-axis direction, there is uncertainty in both the location and how it changes with temperature. The position is likely to be pinned by the strain fields introduced by the FIB milling process. We explored the pinning effect on the antiferromagnetic domain inside $Sr_2IrO_4$ sample through thermal cycling experiments. As seen in Fig. 4(d), only one antiferromagnetic domain is formed after thermal cycling, but its shape appears to be slightly rearranged, which suggests that the antiferromagnetic domain is pinned to structural defects in the crystal. However, this could be due to competition between multiple twin domains, resulting in different profiles of the visible antiferromagnetic domain.

Due to the instability of the MMR stage and the limited flux of the third-generation coherent X-ray beam, we consider the current temperature study to be at its limits, based on many repeated attempts to perform the experiment. In the future, with the development of a high-stability cooling stage and upgrading of the APS storage ring and beamline, we hope to use the resonant BCDI method to investigate the transition of the antiferromagnetic domain in much greater detail to provide insight about the evolution of antiferromagnetic domains near the phase transition. Meanwhile, on a larger length scale, scanning x-ray imaging studies have shown the coexistence of domains with antiferromagnetic spin components along the a- and b-axes in the same crystal, separated by hundreds of microns [6]. Since the isolated $Sr_2IrO_4$ crystal size can be well-controlled by the FIB milling process, this method will give us a future opportunity to further utilize multiple crystal and antiferromagnetic peaks to study the effect of the intrinsic strain on the formation of the antiferromagnetic domains. Typically, the damage of sample from FIB is ~20 nm with 30-keV ion beam and can be further reduced with lower energy beam. Although a 1-keV ion beam was used for final treatment of our samples to minimize the damage, some of the strains and non-sharp edges could be partly caused by the FIB-milling process.

In conclusion, the resonant x-ray magnetic BCDI method has been applied to study the 3D antiferromagnetic domains inside a small crystal of the spin-orbit Mott insulator $Sr_2IrO_4$. The results show a distinct anisotropy of the antiferromagnetic domain between the in-plane and c-axis directions. Using the obtained real-space images of the antiferromagnetic domain, the detailed morphological evolution of the antiferromagnetic domain inside the crystal has been studied. Upon increasing the temperature, the single antiferromagnetic domain observed shows an anisotropic change in its domain size. A possible reason for this observation is due to the magnetic anisotropy of strontium iridate, arising from its strong spin-orbit coupling and crystal field effect. Our study demonstrates that by using resonance, BCDI can be extended to the purely magnetic diffraction signals arising from antiferromagnetic ordering. With beamline and coherent source improvements, even finer details of magnetic structure could be investigated inside such a crystal. A quantitative analysis of the structure of the antiferromagnetic domains within a single crystal can provide important clues about the lattice spin and charge coupling associated with magnetism. The presented approach will find important applications in systems where complex phases give rise to novel physics. We have demonstrated a way to visualize in 3D the antiferromagnetic domains inside an antiferromagnetic material.

**Acknowledgments**

The work at Brookhaven National Laboratory (BNL) was supported by the U.S. Department of Energy (DOE), Office of Basic Energy Sciences (BES), under Contract No. DESC0012704. The sample was prepared using Focused Ion Beam at the Center for Function Nanomaterials at BNL, which is supported by DOE, BES Users Facility Division. The work at UCL was supported by EPSRC. The measurements were carried out at the Advanced Photon Source (APS) beamline 34-ID-C, which was supported by the DOE, Office of Science, BES, under Contract No. DE-AC02-06CH11357. The beamline 34-ID-C was built with U.S. National Science Foundation grant DMR-9724294. T. A. A. would like to acknowledge support by the DOE, Office of Science, BES, Materials Sciences, and Engineering Division, under Contract DE-AC02-76SF00515. G. C. and H. Z. acknowledge support by NSF via Grants No. DMR 1903888 and DMR 2204811.

*lwu@bnl.gov
†irobinson@bnl.gov

antiferromagnetic Bragg reflections from the ~8 um sample. The Supplemental Material also contains Refs. [32-37].

Supplemental Materials for

# Anisotropy of Antiferromagnetic Domains in a Spin-orbit Mott Insulator

Longlong Wu[1*], Wei Wang[1], Tadesse A. Assefa[1,2], Ana F. Suzana[1], Jiecheng Diao[3], Hengdi Zhao[4], Gang Cao[4], Ross J. Harder[5], Wonsuk Cha[5], Kim Kisslinger[6], Mark P. M. Dean[1] and Ian K. Robinson[1,3†]

[1]*Condensed Matter Physics and Materials Science Department, Brookhaven National Laboratory, Upton, NY 11973, USA*

[2]*Stanford Institute for Materials and Energy Sciences, SLAC National Accelerator Laboratory, Menlo Park, California 94025, USA*

[3]*London Centre for Nanotechnology, University College London, London, WC1E 6BT, United Kingdom.*

[4]*Department of Physics, University of Colorado at Boulder, Boulder, Colorado 80309, USA*

[5]*Advanced Photon Source, Argonne, Illinois 60439, USA*

[6]*Center for Functional Nanomaterials, Brookhaven National Laboratory, Upton, New York 11793, USA*

---

[*]lwu@bnl.gov, [†]irobinson@bnl.gov

**Supplemental Figures:**

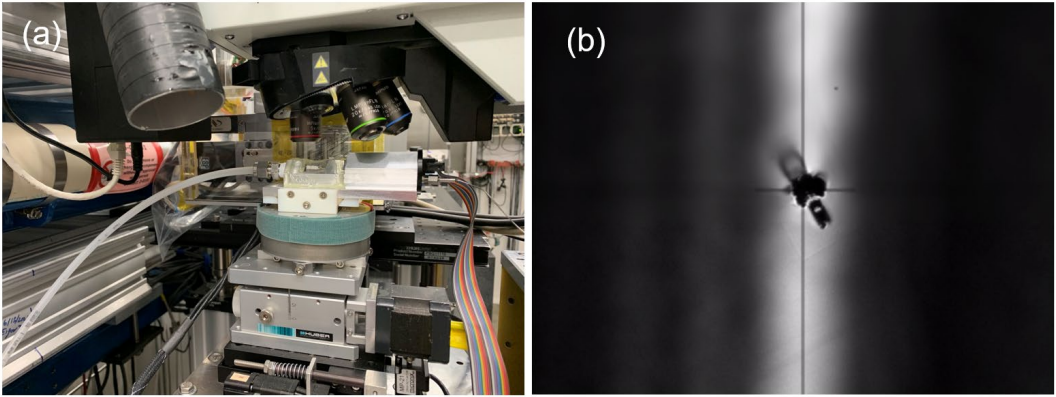

**FIG. S1** Photographs of the experimental setup for the resonant Bragg coherent x-ray diffraction experiment. (a) Picture of the micro miniature refrigerator (MMR) stage during the BCDI experiment. (b) $Sr_2IrO_4$ sample under confocal microscopy at the 34-ID-C beamline.

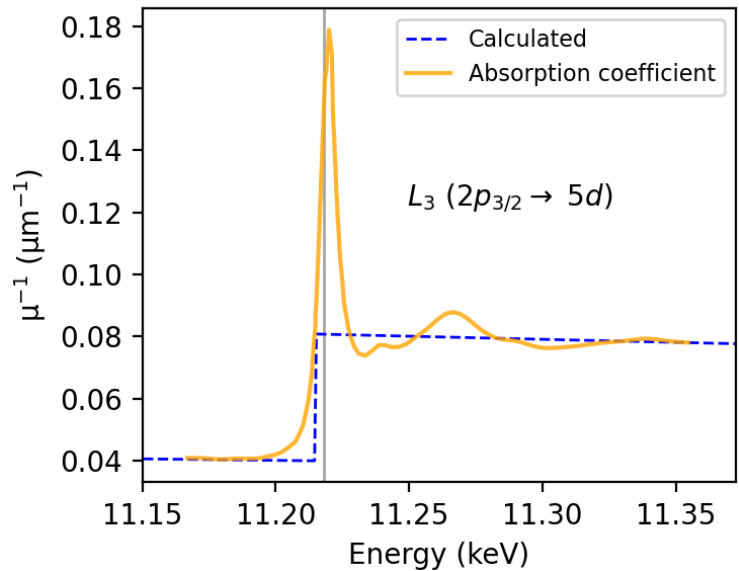

**FIG. S2** Estimated x-ray penetration depth for the $Sr_2IrO_4$ sample. The estimation was completed by scaling the pre-edge and post-edge regions of the x-ray absorption coefficient with the calculated one as labeled by the blue dashed line [1,2]. The corresponding experimental x-ray absorption spectrum is obtained from Ref. [3]. Here, the grey line shows the position of the resonance (11.218 keV), which is found to maximize the intensity of the resonant x-ray magnetic scattering from the sample.

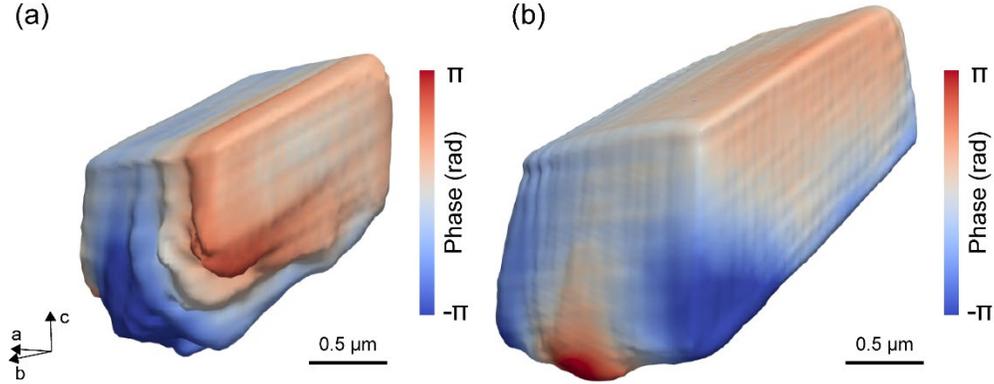

**FIG. S3** Comparison of reconstructed $Sr_2IrO_4$ structure from two different samples prepared by Focused Ion Beam (FIB) milling method, separately. (a) Isosurface of the $Sr_2IrO_4$ sample reconstructed from the (116) peak of the crystal, as the corresponding SEM image shown in Fig 1. (b) Isosurface of another $Sr_2IrO_4$ sample from the (116) peak. Here the two samples were prepared with the same method but with slightly different sizes. For both crystals, near the bottom, a larger strain is shown, which is probably induced by a larger beam current (i.e., ~2.4 nA) for the bulk milling of the $Sr_2IrO_4$ sample (see FIB Sample Preparation section for details). After bulk milling, the bottom region is welded directly to the Pt and its larger strain is retained, while the top and sides were polished with low beam current.

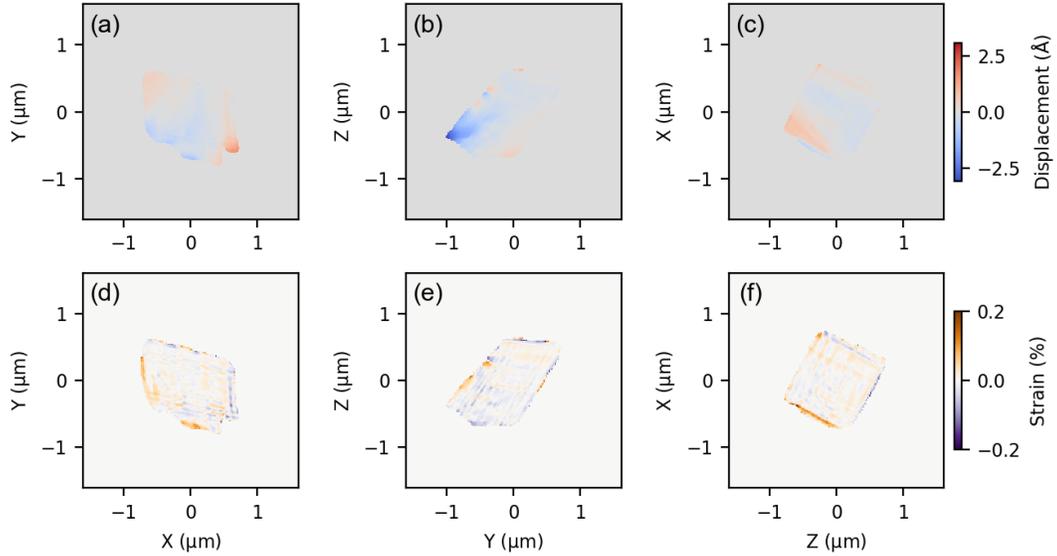

**FIG. S4** Calculated strain inside the $Sr_2IrO_4$ crystal by using the (116) peak. Central slice of the displacement distribution (a) perpendicular to the **Q** vector [i.e., (116) peak]. (b)-(c) along the **Q** vector. (e-f) Corresponding strain distribution.

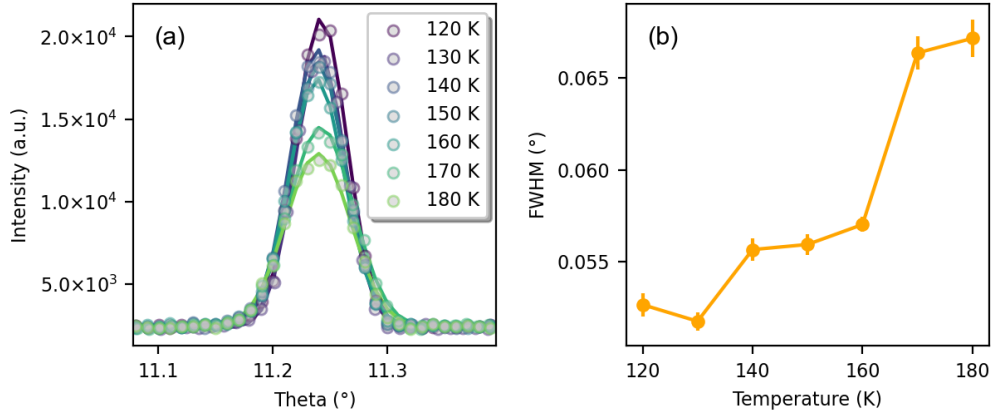

**FIG. S5** (a) Representative rocking curves of antiferromagnetic (106) Bragg reflection measured from an ~8 μm $Sr_2IrO_4$ crystal as a function of the crystal temperature. Here, the dots represent the experimental data, and the lines represent the corresponding fitted results with Pseudo-Voigt function. (b) Extracted full width at half maximum (FWHM) of the antiferromagnetic (106) peak from (a).

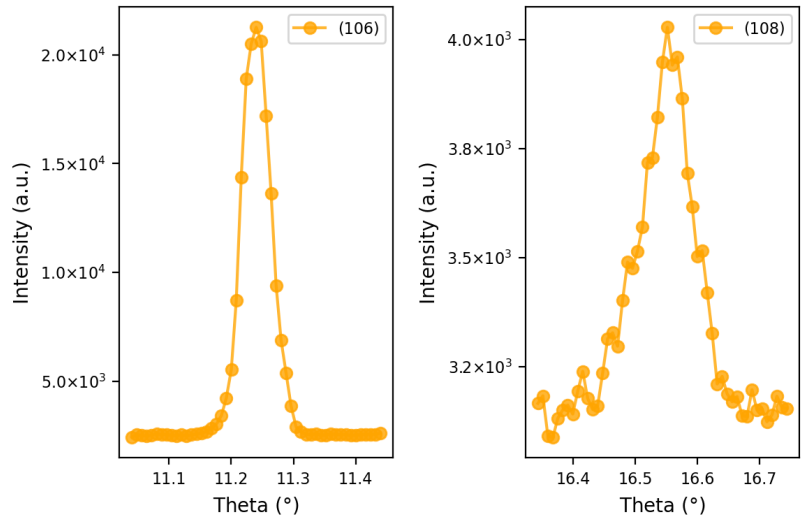

**FIG. S6** Representative rocking curves of antiferromagnetic Bragg reflections at (106) and (108) from an ~8 μm $Sr_2IrO_4$ single crystal prepared by the FIB lift-out method. Both measurements are obtained at 120 K.

## Supplemental Methods:

### Sr$_2$IrO$_4$ Crystal Growth:

The high-quality Sr$_2$IrO$_4$ single crystal was grown from off-stoichiometric quantities of SrCl$_2$, SrCO$_3$, and IrO$_2$ using self-flux techniques [4]. The corresponding molar ratio for the materials is 1.8:1.0:15. The mixture was melted at 1300 °C and subsequently cooled down to 900 °C at a rate of 8 °C per hour before being furnace-cooled to room temperature.

### FIB Sample Preparation:

After the crystal was grown, to prepare a micro-size Sr$_2$IrO$_4$ crystal with a well-defined shape for Bragg coherent x-ray diffraction experiments, the large Sr$_2$IrO$_4$ crystal was pre-oriented crystallographically using a Laue diffractometer. After alignment, its [001] crystal direction was along the vertical direction and the [110] crystal direction was along the horizontal direction. The tetragonal unit cell with a = b = 5.49 Å and c = 25.8 Å at room temperature was used to index the reciprocal space.

During the sample preparation using Focused Ion Beam (FIB) method, the FIB lift-out protocol was then used to cut a micro-size Sr$_2$IrO$_4$ sample from the corresponding large crystal [5]. First, a 1 μm thick sacrificial platinum cap was deposited over the implanted surface of the Sr$_2$IrO$_4$ by electron beam assisted deposition. After the platinum deposition, a lift-out sample, similar to that used for TEM or atom probe tomography, was prepared using a Helios G5 Dual Beam SEM/FIB Microscope. A 30 keV Ga ion beam with 2.4 nA and 9.1 nA beam current was applied for the bulk milling of the Sr$_2$IrO$_4$ sample. After this, a ~5×5×5 μm$^3$ sized sample was extracted/lifted from the bulk using Omniprobe needle, and the cube sample was attached to a 1×1 cm$^2$ silicon wafer using a platinum weld. Then, the lifted Sr$_2$IrO$_4$ sample was trimmed down to a size of approximately 1.2 µm in each direction using lower beam voltages (30 keV, 16 keV, 8 keV, and 5 keV) and smaller beam currents (0.26 nA, 0.12 nA, 63 pA, and 61 pA). Finally, one keV low-energy final cleaning was used to clean off the surface damage from the previous FIB milling step. Such preparation eliminates most of the damage from previous FIB milling steps.

**BCDI Experiments:**

The Bragg coherent x-ray diffraction experiments were carried out at the 34-ID-C beamline of the Advanced Photon Source. A double crystal monochromator was used to select the energy of 9.0 keV and Kirkpatrick–Baez mirrors were used to focus the beam to $600 \times 600$ nm$^2$. To fully illuminate the crystal for valid Bragg coherent x-ray imaging, the beam size was adjusted to $2.2 \times 1.4$ µm$^2$ (H×V) by changing the coherence-defining entrance x-ray slit to $10 \times 20$ µm$^2$ (H×V). The sample was mounted on a micro miniature refrigerator (MMR) stage for BCDI measurements, where the temperature of the sample was controlled precisely by an MMR K-20 controller under a fixed flow of nitrogen gas. Since the Sr$_2$IrO$_4$ sample was pre-aligned before FIB preparation, the precise crystal alignment was quickly determined by using the (204) and (116) peaks of the sample, with a six-axis diffractometer to maneuver the sample orientation. During alignment, the energy of the incident x-ray was 9 keV. After the Sr$_2$IrO$_4$ crystal was well aligned, the energy of the incident x-ray beam was tuned to 11.218 keV, which was found to maximize the intensity of the resonant x-ray magnetic scattering. During the measurements, the resonant x-ray magnetic BCDI data were collected by rocking the sample around the (116) and (106) peaks, separately. The corresponding diffraction signal was collected by a Timepix photon-counting detector mounted 2.5 m away from the Sr$_2$IrO$_4$ sample, and a vacuum flight tube was used to avoid air scattering contribution in the diffraction signal.

**BCDI Data Reconstruction:**

Before feeding the Bragg coherent x-ray diffraction data into a phase retrieval algorithm [6-8] developed in Python, the white-field correction, dark background removal, and hot pixel removal were applied. When applying the phase retrieval algorithm, the measured Bragg 3D diffraction patterns (in detector coordinates) were used as input to the iterative phase-retrieval scheme to reconstruct their corresponding real-space structure information. During the reconstruction, the initial support size of the particle in real space was half the size of the input diffraction pattern array in each dimension. The algorithm was started with 200 steps of relaxed averaged alternating reflection (RAAR). Then, it was switched between hybrid input-output (HIO) with $\beta$=0.9 and error reduction (ER) after every 50 iterations. After the first 100 iterations, the shrink-wrap method was applied in real space to dynamically update the support every ten iterations [9]. At the end of the reconstruction, 200 steps of error reduction were used. The total number of iterations was 1800. After the reconstruction, all the reconstructed results were converted from the detector to sample coordinates. The computation was performed on a computer with 196 GB of RAM and two NVIDIA RTX A5000 GPUs.

**Lattice Strain Calculation:**

For a single Bragg peak, there is a simple linear relationship between the observed image $\phi(\mathbf{r})$ phase and the crystal displacement field $\mathbf{u}(\mathbf{r})$: $\phi(\mathbf{r}) = \mathbf{Q} \cdot \mathbf{u}(\mathbf{r})$ [10]. The strain is related to the variation of the *d*-spacing of the Sr$_2$IrO$_4$ crystal based on the measured Bragg peak. For the Sr$_2$IrO$_4$ crystal (116) peaks, the strain can be calculated as: $\epsilon_{106} = \frac{\partial u_{106}}{\partial x_{106}}$, where $u_{106}$ is the corresponding displacement field.